# Emission Properties of a Hybrid Metallized Diamond(001) Photocathode

L. A. Angeloni, I-J. Shan, and W. Andreas Schroeder*
Department of Physics, University of Illinois Chicago, Chicago, IL 60607-7059, USA

**ABSTRACT**. The spectral emission characteristics of a proof-of-concept hybrid metallized diamond(001) photocathode are presented. The quantum efficiency (QE) is shown to be determined by the photo-injection efficiency across the ohmic contact at the back metallized face, whereas the mean transverse energy (MTE) of the photoemitted electrons is consistent with the a (optical)phonon-assisted and momentum-resonant Franck-Condon mechanism [*Phys. Rev. Applied* **23**, 054065 (2025)] following electron drift transport to the untreated diamond(001) front emission face. Emission is observed from both the lower and upper conduction bands of diamond with positive and negative electron affinity respectively. A potential route to the realization of a red-visible photocathode based on this concept with a sub-50meV MTE and a QE greater than 0.1% is discussed.

## I. INTRODUCTION.

High-brightness and chemically robust solid-state laser-pulse-driven photocathodes that emit electrons at visible wavelengths are highly sought after as front-end electron sources [1-4] for x-ray free-electron lasers (XFELs) [5,6], ultrafast electron diffraction (UED) systems [7-11], and single-shot transmission electron microscopes [12-14]. Several semiconductor materials have been investigated for (and employed in) these applications; for example, $Cs_2Te$ [15-18], $Cs_3Sb$ [19-23], the bi-alkali antimonides [24-28], and Cesiated semiconductors [29-35]. Most (if not all) of these high-brightness photocathodes are not sufficiently chemically robust to survive in air [1,2,36,37] and so require a (ultra)high vacuum environment for reliable operation and vacuum suitcases for transport. In contrast, robust ultra-wide bandgap semiconductors like diamond [38,39] and Gallium oxide [40] have not generally been considered for such applications, primarily due to their large intrinsic ~6eV work functions (valence band maximum to vacuum level). Fortunately, recent successful developments in Ohmic metal contacts for wide bandgap semiconductors [41-44] together with *n*-type doping provides an opportunity for a hybrid photocathode design in which the alignment of the Fermi level across the back metal-semiconductor interface allows for photo-injection of electrons from the metal into the conduction band (CB) of the semiconductor photocathode at visible to near ultraviolet wavelengths ($\lambda \lesssim 350$nm). The subsequent cooling and transport dynamics of the photo-injected electrons in the semiconductor's CB determine their temperature $T_e$ at the photocathode's front emission face and hence, with the electron affinity $\chi$ of the surface, the photocathode's emission properties [45-47].

In this article, we characterize the spectral emission characteristics of a proof-of-concept hybrid metallized diamond photocathode that exploits the recent development of Ohmic metallic contacts for diamond [41] to produce, using *n*-type Nitrogen doping, a robust photocathode with a sub-3.5eV near ultraviolet (UV) photoemission threshold. Similar diamond-based electronic structures with *p*-type doping have been demonstrated for both 'solar blind' UV [48] and visible [49] photodetectors. For our photoemission device, both the quantum efficiency (QE) and the mean transverse energy (MTE) of electron photoemission are measured as a function of the incident photon energy – the latter being defined as $MTE = (\Delta p_T)^2/2m_0$, where $\Delta p_T$ is the two-dimensional root-mean-square (rms) transverse momentum of the emitted electrons (i.e., parallel to the photocathode surface) and $m_0$ is the free electron mass. The MTE of the detected electron emission from the planar diamond(001) crystal face is shown to be dominated by a phonon-mediated Franck-Condon (FC) mechanism [47] due to the strong optical deformation potential scattering in diamond [39,50]. Emission from both the lower and upper conduction bands of diamond are observed. We also discuss approaches to improve the performance of hybrid metallized diamond photocathodes including the possibility of obtaining a red-visible photocathode with a sub-50meV MTE and a QE greater than 0.1%.

## II. EXPERIMENT AND ANALYSIS DETAILS

The studied metallized single-crystal diamond(001) photocathode was fabricated by USA Applied Diamond Inc. [51]. The 5×5mm faced cuboid of 500μm length along the (001) emission direction was homoepitaxially grown via microwave assisted chemical vapor deposition (CVD). A further 10μm of 10ppm nitrogen doped material was grown on one of the faces (the "back" face) of the sample to reliably pin the Fermi level at the upper nitrogen state ~2.4eV below the conduction band minimum (CBM) [38,52,53] (Figure 1). On this doped face, 100nm of titanium, 100nm of platinum, and 1000nm of gold were sequentially deposited, resulting in the formation of a titanium-carbide interface layer (an 'ohmic' contact) between the polycrystalline metals and the diamond after thermal post-deposition annealing [41]. The opposite face (the emission or "front" face) was polished to an optical quality with less than 2nm Ra, so that surface roughness effects [54,55] are not expected to influence the spectral characterization experiments to any significant extent.

*Contact author: andreas@uic.edu

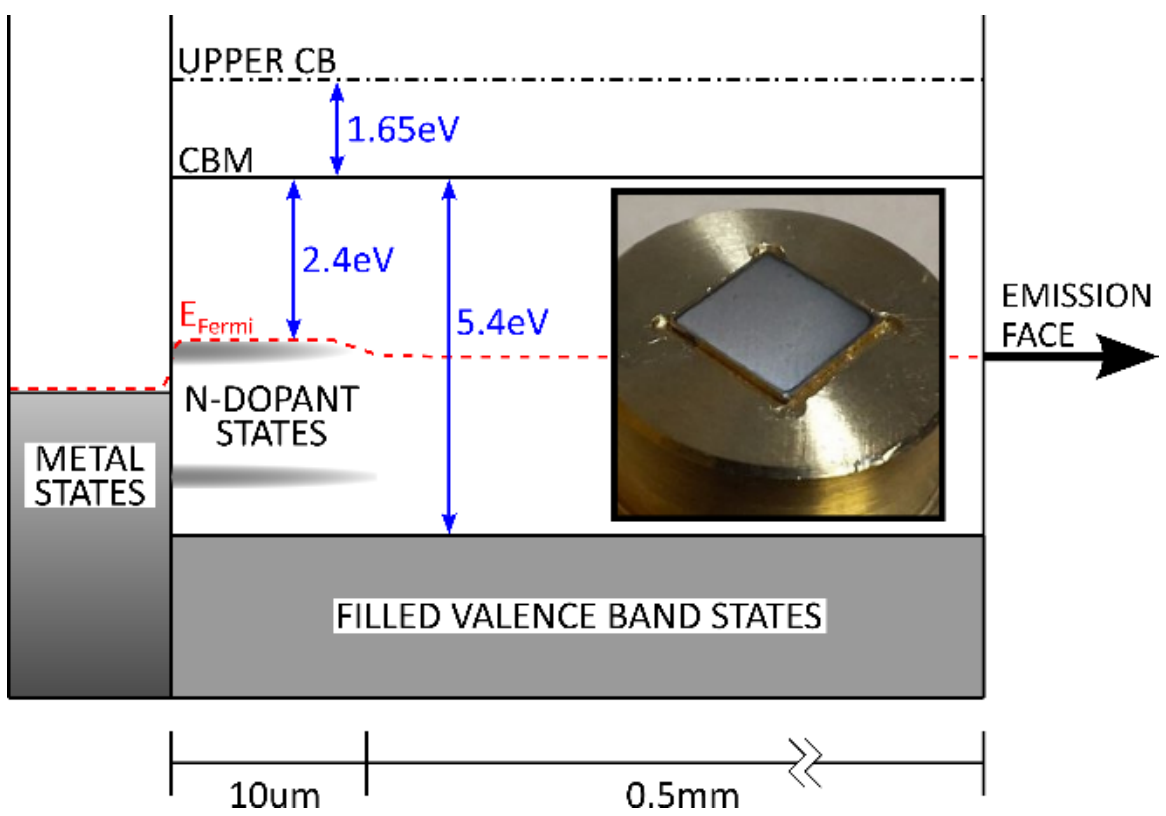

FIG 1. Energy-space schematic of the hybrid metallized diamond(001) photocathode. A 10µm Nitrogen-doped diamond layer pins the Fermi level ($E_{Fermi}$) at the 2.4eV activation energy of the upper Nitrogen dopant state, pulling the Ti metal states' Fermi level to its energetic vicinity. Photo-injection of electrons from the metal states into the upper and lower conduction bands (separated by 1.65eV) provide the source for emission from the photocathode face after transport across its 0.5mm thickness. The inset shows the mounted 5×5mm, 0.5mm-thick metallized diamond(001) photocathode.

The experimental QE and MTE data was obtained using the sub-picosecond tunable UV laser and the 10-20kV DC gun-based, 300K, photocathode characterization system described in Refs. 56 and 57. Briefly, a 20W, 16.7MHz femtosecond Yb:fiber laser system drives the optical parametric amplification of a nonlinear fiber generated continuum which is upconverted by sum frequency generation to produce ~0.5ps UV pulses. The resultant 10-100µW, *p*-polarized, 230-400nm radiation is focused at a 60° angle of incidence to an area of $2\times10^{-4}$cm$^2$ (~120×240µm elliptical spot size) on the photocathode surface producing an incident pulse intensity of less than 100kW/cm$^2$ – an irradiance that is much too small for significant two-photon absorption in diamond [58]. We also note that our entire 3.1-5.3eV UV tuning range is below the 5.4eV indirect bandgap of diamond [59,60], implying that all photoemitted electrons originate from either the polycrystalline Ti metal deposited on the back photocathode face or the N-dopant states within 10µm of the back face. As the occupied density of states for the Ti metal is far greater than that of the nitrogen dopants, the former can be expected to dominate. After photo-injection across the Ti-diamond interface, the electrons are transported in the diamond conduction band(s) to the hybrid photocathode's front emission face (Figure 1) as a result of the internal field in the diamond generated by the applied ~1MV/m DC gun field. After acceleration in the DC gun, the photoemitted electrons enter a 42cm drift region before detection using a micro-channel plate (MCP)/phosphor screen detector (Beam Imaging Solutions; BOS 18) the output of which is imaged with 5:8 demagnification onto a 2.5µm pixel CMOS digital camera (Axiom Optics; FL-20BW), producing a ~20µm optical response point spread function corresponding to a ~1meV MTE resolution. The photocathode characterization system is calibrated for MTE measurements using accurate electron trajectory simulations and for QE measurements using the calibrated photocurrent detected by a Faraday cup from a Rh(110) photocathode [56].

Prior to the photocathode characterization measurements, the undoped front surface is irradiated for typically 20-30 minutes at ~1W/cm$^2$ by the 257nm ($\hbar\omega$ = 4.81eV) fourth harmonic of the Yb:fiber laser system. This *in-situ* 'laser cleaning' step in the ~$10^{-8}$ torr vacuum chamber removes any organic surface contaminants, including those from an initial optical cleaning of the photocathode with propanol, and is performed with the electron gun under a normal 12-16kV operating bias. The laser cleaning process is terminated after ~20 minutes once the photocurrent (i.e., QE) stabilizes. It is possible that this sub-bandgap 'hard' UV-C laser irradiation also modifies the electronic properties of the back metal-diamond interface by, for example, redistributing the interface charge states by photoexcitation and recombination, but we expect this to be a minor effect for an ohmic contact.

Figure 2 displays a typical transverse electron beam profile measured by the CMOS digital camera for the hybrid metallized diamond(001) photocathode, in this case obtained at $\hbar\omega$ = 4.45eV. A thin transverse section through the center of this spatially resolved photo-electron signal is readily fit (right panel in Figure 2) by a function of the form $A\exp(-x^2/w_1^2) + B\exp(-x^2/w_2^2) + C$ where, for our calibrated detector [57], the Gaussian widths $w_i$ ($i$ = 1,2) are directly related to the individual rms momenta of the two separated emission signals and $C$ is a background that accounts for relatively uniform light scatter in our 18mm-diameter MCP/phosphor electron detector and any residual electronic readout noise after background subtraction. For all measured spatial beam profiles, the two Gaussian beam decomposition is mathematically robust with (i) the residual of the least squares fit more than an order of magnitude less than that for a single Gaussian (plus constant) fit and (ii) generally $(A + B)/C > 100$. Moreover, at each incident photon energy, the ratio of the extracted beam widths $w_1/w_2$ is independent of DC gun voltage (i.e., overall detected beam size), the employed conservative MCP amplification voltages, and the camera exposure time, providing confidence that neither of the two extracted Gaussian signals are measurement artifacts. However, the uncertainty in this analytical decomposition increases as the widths of the two Gaussians (i.e., the MTEs of the two photo-electron emission sources) become more equal to each and their amplitudes (i.e., relative QEs) diverge from each other. Nonetheless, the evident clean separation of the two signals in this manner allows their spectral emission characteristics to be studied independently. As discussed below, the spectral QE and MTE characteristics of these two beams are consistent with the emission of electrons populating the upper and lower conduction bands of diamond [38].

*Contact author: andreas@uic.edu

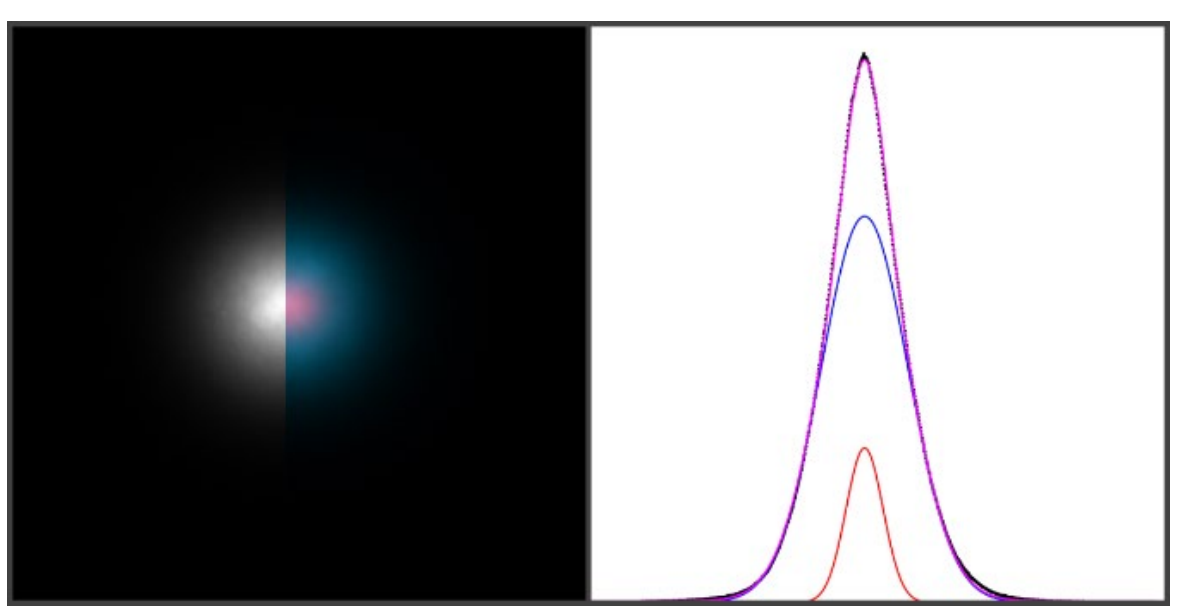

FIG 2. Two-component Gaussian signal decomposition. Left panel: Raw background subtracted digital signal at ħω = 4.45eV (left half) and false color image of the two signal components (right half). Right panel: Fitted Gaussian spatial beam profiles to a horizontal section through the peak of the digital data (black dots). The sum of the weaker inner (red) and stronger outer (blue) Gaussians generate the fit (pink). The ~3:1 ratio of the outer to inner beam widths immediately give a ~10:1 MTE ratio as the horizontal (pixel) axis is directly related to the transverse momentum. Together with the ~3:1 peak height difference, this leads to ~30:1 ratio in the outer to inner signal strengths; that is, the smaller inner signal is ~3% of the total signal.

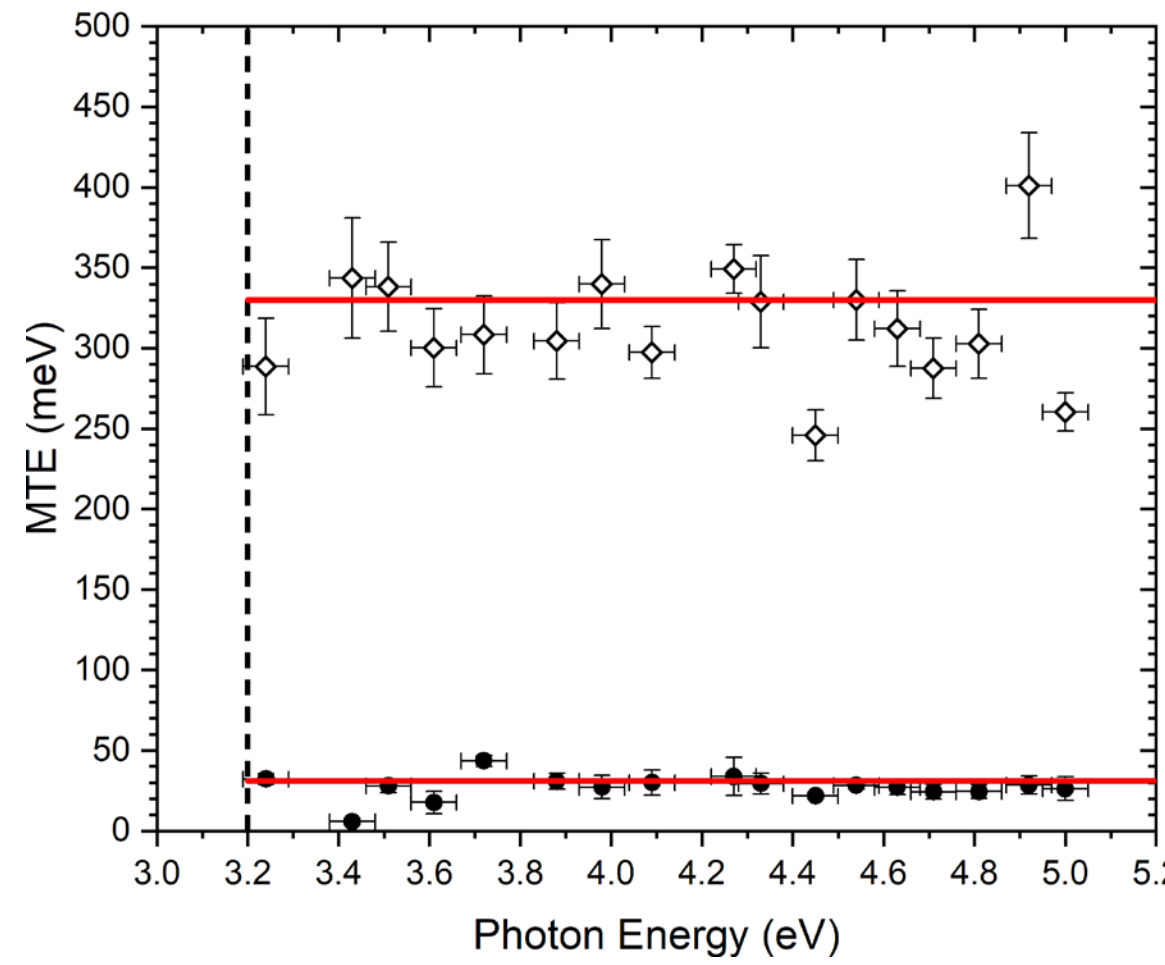

FIG 3. The measured MTE as a function of the incident photon energy for the inner (filled circles) and outer (open squares) signals. The horizontal red lines represent fits to the data using the momentum-resonant phonon-mediated FC emission mechanism (see text). The vertical dashed line corresponds to the observed internal photo-injection threshold of $\phi_{int.}$ = 3.2eV.

## III. EXPERIMENTAL RESULTS AND DISCUSSION

Figure 3 displays the measured spectral dependences of the MTE from both the inner (lower MTE) and outer (higher MTE) beam signals from the 0.5mm-thick metallized diamond(001) photocathode. Immediately evident, after an emission threshold of ~3.2eV, is that (i) the MTE of both signals is constant as a function of the incident photon energy (i.e., independent of ħω) and (ii) the outer signal MTE is ~10× greater than the inner. These two observations are consistent with the 'long transport' regime for semiconductor photocathodes and the emission of electrons via a phonon-mediated, momentum-resonant, Franck-Condon process [39,47]. While the former is dictated by the physics of electron scattering processes [61] in the 0.5mm-thick diamond photocathode, the latter has already been shown to be the dominant emission mechanism for semiconductor photocathodes with a strong optical deformation potential [47] – optical phonon scattering allowing for emitted electrons to be both energy and momentum resonant with the recipient vacuum states. As shown in the Appendix, for a thermalized non-degenerate electron distribution, the MTE (and QE) resulting from this type of emission mechanism is primarily dependent upon the electron affinity χ of the emitting band state and the temperature of the electron distribution $T_e$. For the case of a strong scattering interaction between electrons and optical phonons (i.e., a large dimensionless Fröhlich coupling constant $g$), as is the case for diamond [50], the optical phonon energy ħΩ also influences the photoemission characteristics as multi-phonon scattering becomes more probable during electron emission into the vacuum [39].

*Contact author: andreas@uic.edu

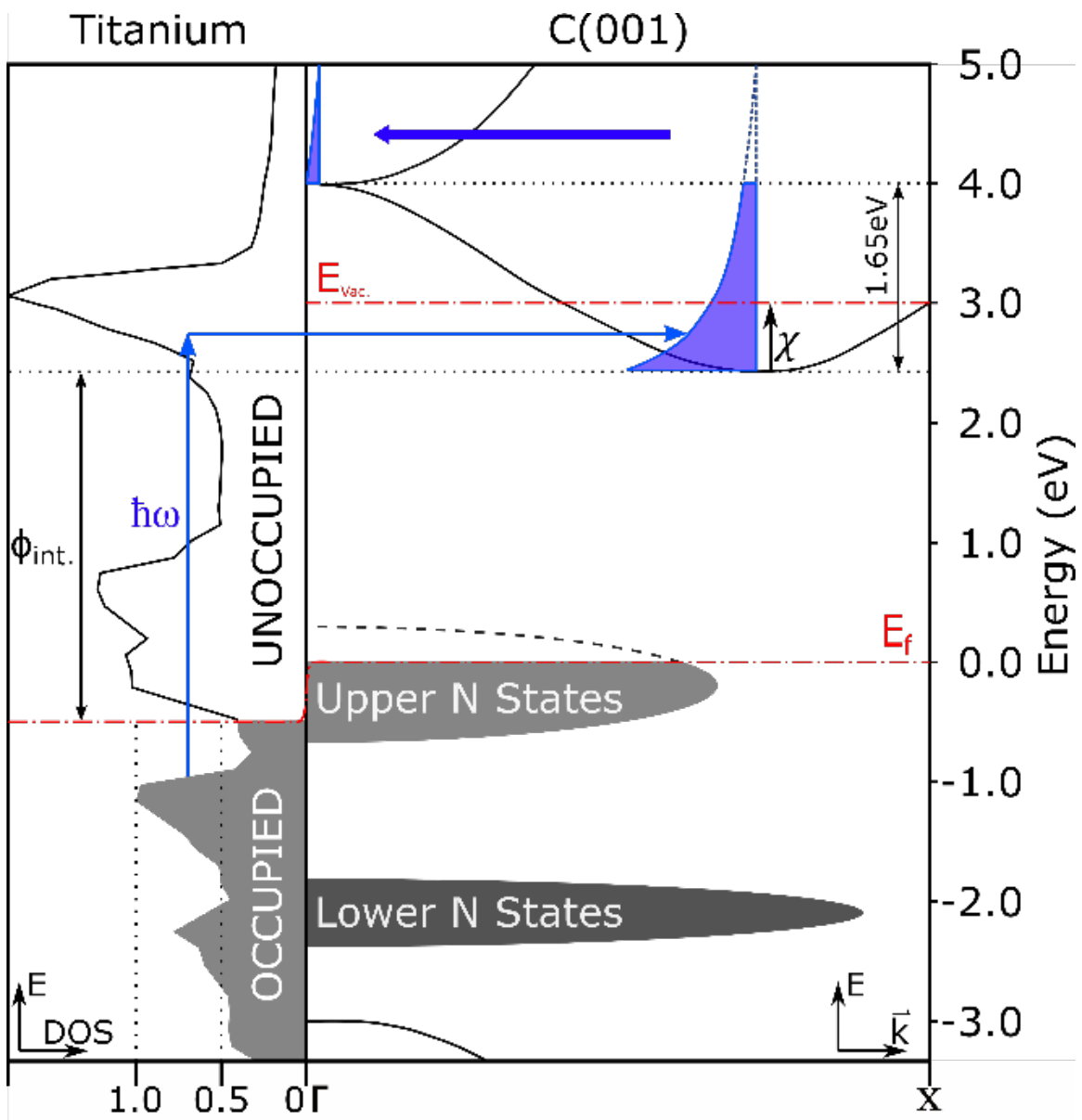

FIG 4. Density of states for polycrystalline Ti metal (left) and the Γ-X band structure of diamond with the upper and lower Nitrogen dopant states within the first 10μm of the metal-diamond interface (right). Electron photo-injected across the $\phi_{int.}$ potential barrier from the occupied metal states into the diamond conduction band generate a non-degenerate thermalized distribution (blue) the high energy portion of which populates the upper conduction band. For a bare, untreated diamond(001) surface, the vacuum energy level ($E_{vac.}$) resides between the lower and upper conduction band minima (as shown).

A key to understanding the spectral emission properties of this hybrid diamond photocathode is therefore the determination of the temperature $T_e$ of the electron distribution that drifts towards the (001) crystal emission face under the influence of the applied gun field. After photo-injection from the Ti metal into the diamond conduction band at the metallized back face, the electrons are expected to rapidly cool and generate a thermalized distribution at near the 300K lattice temperature by optical phonon scattering driven by the strong optical deformation potential of diamond [50] irrespective of their initial photo-excitation energy within the conduction bands. Thereafter, the 16kV/cm gun field forces them to drift towards photocathode emission face as the DC dielectric constant of 5.3 for diamond [62] implies an internal field of about 3kV/cm. For single-crystal undoped diamond, electron mobilities of as high as 4,000cm$^2$/(V.s) have been measured at 300K [63]; however, experimental [64] and theoretical [65] studies generally indicate values in the range 2,000-2,500cm$^2$/(V.s), which for the internal applied gun field gives a drift velocity $v_d$ of ~6x10$^6$cm/s for a thermalized electron distribution. This value is about half the measured saturation drift velocity $v_{ds}$ of 1.2x10$^7$cm/s [64] (associated with a saturation field strength of ~15kV/cm [64,65]) – a value in agreement with an analysis of equilibrium electron drift transport under polar optical scattering [61] that calculates $v_{ds} \approx \sqrt{3\hbar\Omega/(4m^* \coth(\hbar\Omega/2k_BT))}$ where $T$ is the 300K lattice temperature, the optical phonon energy ħΩ = 165meV [39,66], and $m^* = 1.2m_0$ is the electron effective mass of the anisotropic LCB in the (001) crystal direction (the drift direction) [60]. The latter analysis, for $v_d/v_{ds} \approx 0.5$ and when $k_BT/\hbar\Omega = 6.6$, gives $T_e/T$ of about 1.2; in other words, $k_BT_e \approx 30$meV for the drifting thermalized electron distribution.

Using this expected value for the average drift transport energy of the thermalized electron distribution and the analysis presented in the Appendix for phonon-assisted FC electron emission we obtain the upper and lower horizontal lines in Figure 3 using a fit to the MTE constrained by the known material and band structure properties of diamond [60]. First and foremost, it is known that the upper and lower conduction bands of diamond are separated in energy by 1.65eV [59,60,67] as shown in Figure 4 and that for a bare (untreated) diamond(001) face the vacuum level is located between the upper and lower conduction band minima [38,68,69]. This means that the LCB has a positive electron affinity $\chi_{LCB}$ whereas the UCB has a negative electron affinity $\chi_{UCB}$ with a corresponding higher intrinsic QE: And one obtains the energy constraint $\chi_{LCB} - \chi_{UCB} = 1.65eV$. Second, the strong optical deformation scattering in diamond leads to a renormalization of the electron energy in the bands due to the formation of polarons [70]. For an intrinsic dimensionless Fröhlich coupling constant $g_0$ and a band electron effective mass $m^*$, the renormalization energy is given by $\sqrt{m^*/m_0}\, g_0\hbar\Omega$ [70,71] – the real part of the electron self-energy shift below the (conduction) band minimum associated with the polaron quasi-particle [39,72]. Consequently, for the electron population in a band, the effective electron affinity $\chi^* = \chi + \sqrt{m^*/m_0}\, g_0\hbar\Omega$. Combined, these two physical properties give the modified energy constraint

$$\chi^*_{LCB} - \chi^*_{UCB} = 1.65eV + g_0\hbar\Omega\left(\sqrt{\frac{m^*_{LCB}}{m_0}} - \sqrt{\frac{m^*_{UCB}}{m_0}}\right).$$

Thus, since $g_0 \approx 1.0$ for diamond [39], with $m^*_{UCB} = 0.42m_0$ [38] and $m^*_{LCB} = 1.2m_0$ in the (001) propagation

*Contact author: andreas@uic.edu

(and emission) direction [60], we obtain $\chi^*_{LCB} - \chi^*_{UCB} \approx 1.72eV$ for the energetic difference between the electron populations in the upper and lower CBs at $k_BT_e \approx 30$meV. Further, as the effective Fröhlich coupling constant $g = \sqrt{m^*/m_0}\, g_0$ is different for the upper and lower conduction bands, the Poisson statistics governing multi-phonon FC emission from the two bands will be different. Specifically, the strength of the $n^{th}$ phonon FC emission process is given by $e^{-g}g^n/n!$ [39,73], where $n$ = 0,1,2,3…, implying that an appropriately weighted (and terminated) sum is required to evaluate the total QE and hence also the MTE of electron emission as described in the Appendix (equations A6 and A7).

The two horizontal lines in Figure 3 use the known physical parameters of diamond ($g_0$, ħΩ, $m^*_{UCB}$, and $m^*_{LCB}$), $k_BT_e$ = 30meV, and values of $\chi_{LCB}$ = 0.81(±0.05)eV and $\chi_{UCB}$ = −0.84(±0.05)eV in fitting the experimental MTE data to Appendix equations A6 and A7 for NEA and PEA FC emission from the upper and lower conduction bands respectively. The uncertainties quoted for the two electron affinities are reflective of the rms fit to the experimental data. Inclusion of uncertainties in the two-signal extraction etc. yield an overall uncertainty of around ±0.1eV, giving $\chi_{LCB} \approx$ 0.6 to 0.8eV which is consistent with literature values for a bare (untreated) diamond(001) surface [68,69]. As this PEA for the LCB is much greater than $k_BT_e$, only electrons in the high energy (exponential) tail of the band's thermalized electron distribution above the vacuum level can emit. As shown by equation A6 for the FC emission mechanism under PEA conditions, this produces an MTE ≈ $k_BT_e$, supporting the suggestion that the measured inner MTE signal is due to LCB emission. The negative value for $\chi_{UCB}$ is then also consistent with the observation of a larger 330meV MTE for the outer signal originating from UCB FC emission as the leading term in equation A7 is proportional to $\frac{1}{2}|\chi_{UCB}|$ = 0.42(±0.05)eV with multi-phonon emissions ensuring a somewhat lower overall MTE due to energy-momentum restrictions in the available vacuum states (see Appendix) [39,47].

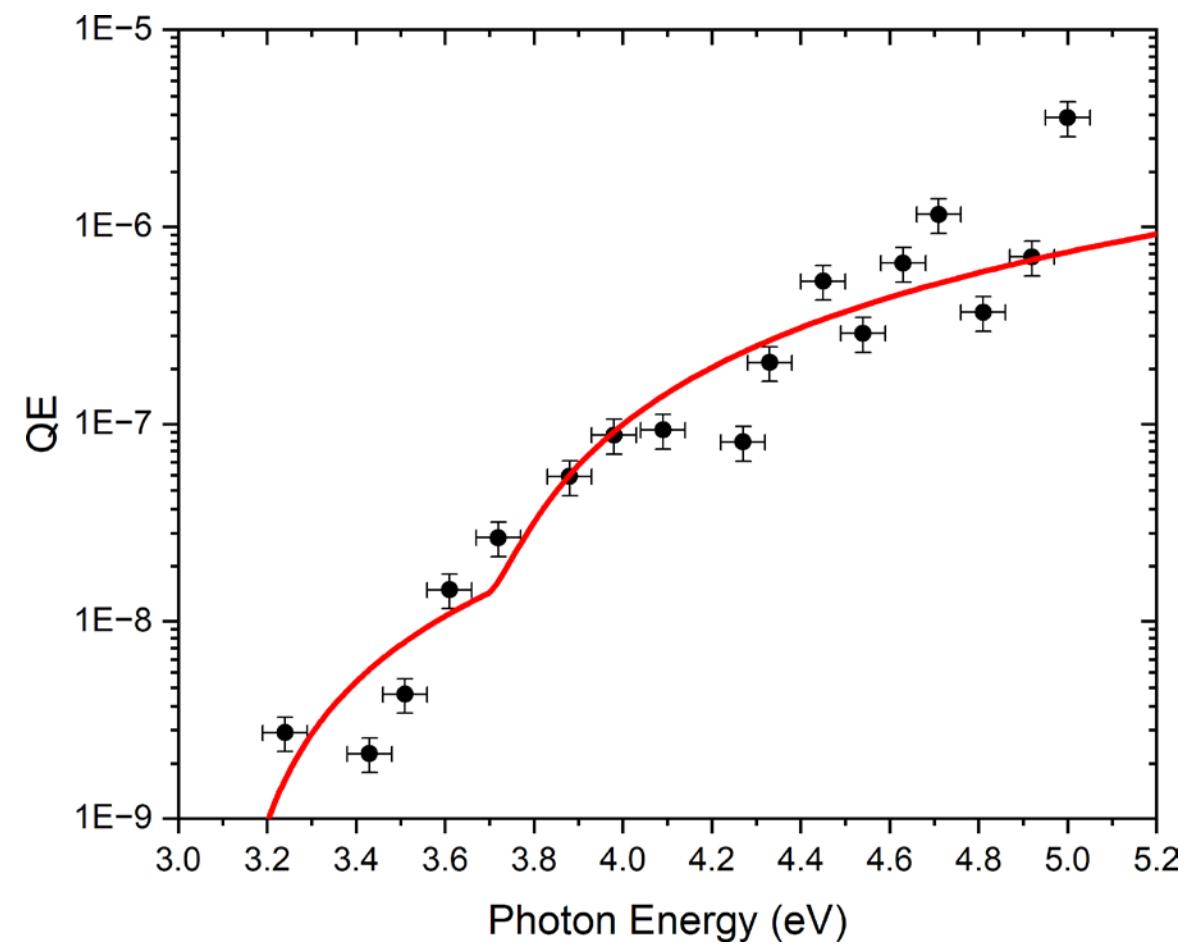

FIG 5. The measured QE as a function of the incident photon energy for the hybrid photocathode. The solid line is a fit to the data of the form $QE = A_1(\hbar\omega - \phi_1)^{3/2} + A_2(\hbar\omega - \phi_2)^{3/2}$, with $\phi_1 = \phi_{int.}$ = 3.2eV and $\phi_2$ = 3.7eV, associated with a two-step increase in the polycrystalline Ti density of states (see Figure 4), and $A_1 = 4\times10^{-8}$eV$^{-3/2}$ and $A_2 = 6\times10^{-7}$eV$^{-3/2}$.

In contrast to the MTE, the spectral dependence of the total QE (the sum of LCB and UCB emissions) is not dependent on the dynamics of the drift electron transport in the hybrid diamond photocathode as for any incident photon energy the electron temperature $T_e$ at the photocathode emission face is constant at about 360K – the parameter that dictates the QE for the FC emission mechanism for the fixed material constants $g_0$, χ, and ħΩ (see Appendix). Consequently, the QE should be solely determined by the injection efficiency across the back ohmic metal-diamond interface as recombination effects during drift transport are insignificant in the high-quality CVD-grown undoped diamond due to filled valence band states. Shown in Figure 5 is the measured spectral dependence of the photocathode's QE together with a fit of the form $QE = A_1(\hbar\omega - \phi_1)^{3/2} + A_2(\hbar\omega - \phi_2)^{3/2}$, where $\phi_1$ = 3.2eV (the emission threshold energy associated with the Fermi level in the Ti metal (see Figure 4)) and $\phi_2$ = 3.7eV which corresponds to the threshold energy required to promote electrons into the LCB from the second step in the occupied density of states (DOS) in the polycrystalline Ti metal evident in Figure 4. The fit therefore assumes a stepwise constant DOS in the Ti metal (the electron source), and a uniform transmission (for each DOS step) over the internal potential barrier into the recipient LCB states described using a standard DOS for a parabolic band of the form $\sqrt{E}dE$ for the energy interval $E$ to $E$+d$E$ integration over which gives the $E^{3/2}$ dependence for the total number of injected electrons. However, the ratio $A_1$:$A_2$ of 1:15 obtained from the fit is significantly greater than the 1:2 ratio in the DOS step (Figure 4). We attribute this difference to the fact that our *ab initio* band structure calculations indicate that the Ti states contributing to the second step in the DOS are generally at momenta that are a better match to the states around the minimum of the LCB, resulting in an increased injection efficiency. The reasonable fit to the experimental data is therefore direct evidence that the QE is only dependent upon the injection efficiency across its back metal-diamond interface (i.e., the vast majority of the emitted electrons originate from the Ti metal), and thus that this hybrid photocathode operates in the long transport regime with $T_e$ just above the lattice temperature [47].

*Contact author: andreas@uic.edu

As the two signals from the upper and lower conduction bands were separated in our analysis, the spectral dependence of the ratio of the UCB to LCB emission quantum efficiencies can also be determined. Within experimental uncertainty of a few percent associated with the two-signal decomposition, the QE of the inner LCB signal is found to be relatively constant at ~5% of the total signal over the entire investigated spectral range. This means that the total MTE of emission from the hybrid photocathode is, to a good approximation, given by that of the outer signal (Figure 3) – the difference is within the uncertainty in the measurements. However, the QE ratio of ~20:1 between the two emission signals favoring the outer UCB signal is in apparent contradiction to the expected theoretical QE ratio for FC emission from the two bands. Specifically, for $k_B T_e$ = 30meV and in the limit of zero phonon energy ($\hbar\Omega \rightarrow 0$), the FC emission mechanism predicts a minimum expected intrinsic QE ratio given approximately by $exp[\chi_{LCB}/(k_B T_e)]$:1 favoring the UCB [47] (see Appendix). On the other hand, the thermalized population density ratio between the lower and upper conduction bands is about $exp[(\chi_{LCB} - \chi_{UCB})/(k_B T_e)]$:1 respectively, giving a net $exp[-\chi_{UCB}/(k_B T_e)]$ ~ $10^{11}$ favoring LCB emission in the employed parabolic band approximation. This discrepancy can be explained by considering factors that the analyses presented in Ref. 47 and the Appendix do not include; in particular, in this case, the real (non-parabolic) energy dependence of the conduction band DOS in diamond and its effect on the scattering rate between electrons and optical phonons [74]. First, in-house *ab initio* band structure calculations indicate that the overall conduction band DOS in diamond increases approximately as $E^2$ rather than the $\sqrt{E}$ dependence of a parabolic band. This means that for a thermalized distribution with $k_B T_e$ = 30meV there are ~$10^3$ more electrons above the UCB minimum (1.65eV above the LCB minimum) than for a purely parabolic conduction band dispersion. Second, the analysis presented in Ref. 74 indicates that the electron-(optical)phonon scattering rate in diamond at 300K is a factor of ~5 greater 1.65eV above the LCB minimum (~$10^{15}$s$^{-1}$) than in the vicinity of its minimum (~$2\times10^{14}$s$^{-1}$). Together these two factors imply that (i) the scattering rate to replenish any emitted electrons, and hence reconstitute the thermalized carrier distribution, in the UCB is at least 1,000× higher than that for the LCB states and (ii) the (optical)phonon-mediated emission rate for the entire population in the UCB is significantly higher than that for the Boltzmann tail of the LCB distribution existing above the vacuum level. The latter is reenforced by fact that for $v_d \approx$ 5-6×$10^6$m/s electrons traverse the expected ~10nm emission region near the diamond(001) surface [47] in only a few femtoseconds, allowing multiple emission attempts for UCB electrons but at most only one for electrons in the LCB. The experimental data indicates that these physical effects sufficiently enhance UCB FC emission so that it becomes ~20× stronger than that from the LCB. A full theoretical understanding of the physics behind FC emission from diamond(001) is expected to require detailed Monte Carlo modeling of electron-(optical)phonon scattering over the entire Brillouin zone.

We note that the above interpretation of the experimental data invokes electronic states (i.e., the bulk band states) that have the largest DOS, even within the ~10nm surface emission region of optical phonon assisted FC emission [47]. Bulk defects and surface states will have a substantially reduced DOS in the surface emission region for our high-quality crystal diamond photocathode. Additionally, in order to emit, such spatially discrete states will need to be above the vacuum level and subject to the same thermal population distribution as the conduction band electrons; that is, a factor of at least $exp[-\chi_{LCB}/(k_B T_e)]$ ~ $10^{-11}$ less than a state near the conduction band minimum. Moreover, for atomic-scale states which have a sufficiently large momentum-space wave function to allow direct momentum-resonant emission into the vacuum, it is straightforward to show that a state (or collection of states) with an energy $E \approx$ 0.5eV above the vacuum level is required to produce the observed 330meV MTE outer signal:

*Contact author: andreas@uic.edu

$$MTE = \frac{\frac{\pi A}{m_0}\int_0^{\pi/2} d\theta sin^3\theta \int_0^{\infty} dp\, p^4\, \delta\left(E-\frac{p^2}{2m_0}\right)}{2\pi A\int_0^{\pi/2} d\theta sin\theta \int_0^{\infty} dp\, p^2\, \delta\left(E-\frac{p^2}{2m_0}\right)} = \frac{2}{3}E.$$

Here the energy conserving delta function selects the vacuum momentum state at energy $E$ and $A$ is a constant that includes the spin degeneracy, mass dependence, etc. of the vacuum DOS. Experimental evidence for such a bulk defect or surface state at around 1.3eV above the conduction band minimum in diamond has, to our knowledge, not been reported.

We also note that both the experimental MTE and QE data clearly indicate that the internal photo-injection threshold $\phi_{int.} = \phi_1 \approx 3.1$eV is around 0.7eV greater than the 2.4eV threshold expected when the Fermi level is pinned by the upper nitrogen dopant level [38,52,53] and energetically level with that in the Ti metal. The likely explanation is the formation of a Ti-C bond dipole at the interface produced by the metallization, plus some downward band bending towards the interface in the doped *n*-type region [42]. The metal-diamond interface is therefore not completely ohmic as it has a directional character akin to a *p*-*n* junction, i.e. a Schottky barrier.

## IV. FUTURE DIRECTIONS

Further optimization of the performance of metallized diamond(001) photocathodes is possible at both the emission face and the back photo-injection interface. For the latter, *n*-type doping with phosphorous rather than nitrogen at the back metallized surface is expected to reduce the photo-injection threshold to less than 1.5eV even with the 0.7eV Ti-C back-surface dipole since the donor activation energy of phosphorous is around 0.57eV [75] giving $\phi_{int.} \approx 1.3$eV with the same Fermi level pinning. This would then produce a 'red' photocathode (photoemission at wavelengths greater than 600nm) with similar spectral properties to those of the N-doped variant characterized above with photon energies shifted down by 1.5-2eV.

The QE of this type of hybrid photocathode can also be dramatically increased by using surface terminations to reduce the work function of the emission face and hence the electron affinity of the LCB where the vast majority of the photoinjected carriers reside in the long transport regime with $k_B T_e \approx$ 30meV. For example, H-termination of a diamond(001) surface has been shown to produce $\chi_{LCB} \approx -0.5$eV [39,76] which would theoretically be expected to increase the front face emission efficiency by a factor of $\exp[\chi/(k_B T_e)] \approx 10^7$ [47], implying that overall QE of the resulting photocathode will likely be solely limited by the back-face photo-injection efficiency. However, this is also expected to produce an MTE of around 280meV (see equation A3 or A7). An emitted electron beam with a smaller MTE of 40-50meV should be generated for near zero electron affinity for the LCB with perhaps only a factor of 10 reduction in the QE (see equation A4 or A5). Provided a suitable robust surface termination can be found to facilitate $\chi_{LCB} \approx 0$ for diamond(001) using a P-doped back face metallization, this could produce a useful 'high brightness' visible (red) photocathode.

In addition, the work of Chang et al. [77] indicates that metallized diamond(001) photocathode could generate an effective QE well above unity through avalanche amplification of the electron emission under high gun fields – an effect that exploits the inherent robustness of diamond. This would, however, likely substantially increase the electron temperature $T_e$ at the photocathode front face and thus the MTE of electron emission. One would also need to be careful in their use in RF guns since the gun frequency $\nu_{RF}$ will need to be significantly less than $v_{ds}/z_{cath.}$ – the reciprocal of the electron transit time over the photocathode length $z_{cath.}$. For the above case, $z_{cath.} = 0.5$mm and using the saturation drift velocity $v_{ds} \approx 1.2\times10^7$cm/s for diamond [64.65] one obtains $\nu_{RF} << 240$MHz; for example, placing the 185.7MHz LCLS-II RF gun [78] too close to the limit unless $z_{cath.}$ is reduced to less than 100μm. However, the smaller photocathode crystal length will also reduce any net avalanche amplification by more than a factor of 100 from $z_{cath.} = 0.5$mm.

We note that there is also a limit to how small $z_{cath.}$ can be if one wishes to maintain a spectrally independent $k_B T_e \approx 30$meV (and hence an MTE < 50meV when $\chi \approx 0$) for diamond in the long transport limit [47]. Specifically, this requires that steady-state drift transport be established. As the characteristic optical photon emission time for an electron in the CB of diamond is ~5fs [74], electrons photo-injected into the CB with an excess energy of 1eV could cool to within $\hbar\Omega = 165$meV [39,66] of the band minimum in around 30fs, but are more likely to initially cool to $k_B T_e \approx$ 200meV – the equipartition electron energy with the two (TO and LO) optical phonon modes [47] – with further cooling delayed by the ~7ps optical phonon decay time [66]. The latter longer time then gives the requirement that $z_{cath.}$ be greater than ~1μm, again using $v_{ds} \approx 1.2\times10^7$cm/s as the limiting electron velocity [64,65]. As this photocathode thickness is less than the 10μm doped region used to pin the Fermi level (Figure 1), which could perhaps be halved, a reasonable practical limit for a sub-50meV MTE hybrid metallized diamond(001) photocathode at 300K is that $z_{cath.} > 10\mu m$. This would give a minimal electron transit (i.e., transport) time over $z_{cath.}$ of 50-100ps and thus allow the operation of a visible (red) photocathode in a sub-GHz RF gun.

*Contact author: andreas@uic.edu

Metallized 'dielectric' photocathodes that employ robust wide band gap semiconductors other than diamond can also be envisioned. For example, ohmic metal contacts on *n*-type (Si-doped) Gallium nitride (GaN) [79] and (Sn-doped) Gallium oxide ($Ga_2O_3$) [80] could potentially allow for efficient photo-injection over a sub-0.5eV back surface potential barrier. The strong optical deformation potential scattering with $\hbar\Omega \approx 90$meV in both materials [81,82] would then be expected to produce $k_BT_e$ less than 50meV for drift transport at 300K [61] and thus a low MTE for electron emission with $\chi \approx 0$ through the described phonon-mediated FC mechanism. Further and in contrast to diamond, the lowest CB in both GaN and $Ga_2O_3$ is located at the Γ point of the Brillouin zone [80,83] which means that direct emission [84] from the CB will be significantly more efficient due to intrinsic near band-to-vacuum momentum matching at $\chi \approx 0$. As these CBs have an electron effective mass $m^*$ of around $0.2m_0$ [80,85], an $MTE \approx \left(\frac{m^*}{m_0}\right) k_B T_e$ of less than 10meV may be expected from direct band emission.

There have also been several prior studies of hybrid dielectric on metal photocathodes where $z_{cath.}$ is significantly less than 100nm [86-92]. In particular, the deposition of a 3nm layer of CsI on Cu(001) resulted in a three orders of magnitude QE enhancement at 266nm [88]. From the viewpoint of this article, these hybrid photocathodes are clearly in the short transport regime (if electrons occupy the dielectric CB states) where equipartition of the electron energy with optical phonon modes and band bending effects are important [47]. The alternative emission mechanism is that the photoemitted electrons are 'transported' through the CB of the thin dielectric layer without appreciable scattering as has been demonstrated for a ~2nm-thick $Cs_2Te$ layer on GaAs [93]. In either case, the MTE of emission at 300K is likely to be large (>100meV) [90] unless (i) the emitting electron distribution has a low temperature ($T_e \rightarrow 300$K) from near threshold photo-injection from the metal [92], and/or (ii) $\chi \geq 0$ if three-dimensional (i.e., near bulk-like) optical phonon scattering effects are prevalent for the confined vibrational lattice modes of the thin dielectric film.

## IV. SUMMARY COMMENTS

The spectral emission properties of a hybrid metallized diamond(001) photocathode are presented and shown to be consistent with a phonon-mediated FC process – a momentum-resonant emission mechanism that is expected to be dominant for semiconductor photocathodes with sufficiently strong optical deformation potential scattering [47]. Two emission signals are observed, both displaying an MTE independent of the incident photon energy; (i) a weak inner signal with a low MTE of around 30meV and (ii) a stronger outer signal with an MTE of 330meV. These two signals can be well described by the FC emission theory (see Appendix) with the appropriate material parameters and the physics associated with drift transport to the diamond(001) emission face. Specifically, a standard semiconductor transport analysis under the applied ~3kV/cm internal field [61] indicates that after rapid initial cooling by optical phonon emission the resultant thermalized electron distribution photo-injected from the metallized back face drifts with $k_BT_e \approx 30$meV over the 0.5mm photocathode thickness; that is, in the 'long' transport regime [47]. This allows the weaker inner signal to be identified as being due to FC emission from the lower conduction band with a PEA of $\chi_{LCB} \approx 0.8$eV, since for FC dominated emission the MTE tends to $k_BT_e$ for $\chi$ much greater than $k_BT_e$ [47] (Appendix equation A1). The larger 330meV MTE of the stronger outer signal may then be rationally associated with FC emission from the upper conduction band with a NEA of $\chi_{UCB} \approx -0.85$eV. This interpretation of the experimental MTE data is further supported by the spectral dependence of the QE which is observed to be consistent with the expected photo-injection efficiency across the back polycrystalline Ti to diamond interface – all the emitted electrons being generated at the hybrid photocathode's back metallized face as the incident photon energies are less than the 5.4eV diamond bandgap, and the constant electron temperature $T_e$ of 360K at all incident photon energies (due to the long transport regime) implying a constant intrinsic QE for emission from both the upper and lower conduction bands.

We note that the agreement between the experimental MTE data and the phonon-mediated FC emission process using just four parameters ($T_e$, $\chi$, $\hbar\Omega$, and the Fröhlich coupling constant $g_0$) is obtained without considering the energy bandwidth of the optical phonons involved in the momentum-resonant scattering of electrons into the vacuum states. In general, one would expect this bandwidth to be convolved with the electron energy distribution in its 'mapping' onto the vacuum states [47], causing the energy width of the emitted electrons to increase as $n$ (the number of phonons involved) increases. Such an effect is expected to be observable when ħΩ is significantly greater than $k_B T_e$ and indeed has been for a hydrogen-terminated diamond(001) photocathode [39]. Its effect on the MTE is however not expected to be significant since the Poisson statistics dictate that the strength of the $n^{th}$ FC emission process is proportional to $e^{-g}g^n/n!$ [39,73] and so declines rapidly as $n$ increases for $n > g$ – noting that $g_0$ is of the order unity for diamond [39].

*Contact author: andreas@uic.edu

It is also to be noted that the results presented here are in apparent contradiction to those of a heavily N-doped photocathode [38] where it was shown that the electron temperature $T_e$ at emission was dependent on the incident photon energy ħω, generating non-monotonic variations in the emitted MTE (and QE) as the average initial electron temperature changed due to different dominant photoexcitation mechanisms. We believe that this fundamental difference in the emission characteristics of the two photocathodes is related to the large difference in their photoexcited electron population densities. For the heavily N-doped case, photoexcitation into the CB states by linear absorption occurs over a 400μm-thick doped region. On the other hand, for the hybrid photocathode, the photo-injected 'sheet' charge is expected to have a thickness much less than 100nm – an upper limit of 100nm being inferred from the product of the $v_{ds} \approx 1.2\times10^7$cm/s saturation drift velocity [64.65] and the ~0.5ps laser pulse duration. For equivalent excitation probabilities into the CB(s) of diamond for our ~10nJ/cm$^2$ UV incident laser pulse fluence, around $10^6$ electrons are excited per pJ-level laser pulse, giving a photoexcited electron density of ~$10^{11}$cm$^{-3}$ for the N-doped photocathode and likely greater than ~$10^{15}$cm$^{-3}$ for the hybrid photocathode. The inter-electron (charge) separation is therefore more than 10× greater for the N-doped case (~1μm) than the metallized diamond(001) photocathode (less than 100nm). Thus, the requirement that $2/g \approx 2$ electrons interact with one optical phonon for polaron-mediated electron transport [61] is much more probable for the hybrid photocathode; indeed, "Frohlich" polarons associated with a long-range electron-phonon interaction have a radius very much greater than the crystal lattice parameter [94] resulting in coherent motion with a free carrier (electron) mobility $\mu_e >> 1$cm$^2$/(V.s) – as is observed in diamond [63]. Conversely, the 'single' electron picture of transport and emission employed to describe the spectral properties of the N-doped diamond(001) photocathode [38] is a also valid approximation.

*Contact author: andreas@uic.edu

**ACKNOWLEDGEMENTS**

The authors are also indebted to the Physical Sciences Machine Shop for their assistance and expertise. This work was supported by the U.S. Department of Energy under Award no. DE-SC0020387.

**DATA AVAILABILITY**

The experimental data presented in this paper is openly available [95].

**APPENDIX**

As outlined in Ref. 47, our theoretical analysis of electron emission by a momentum-resonant phonon-mediated FC mechanism involves the mapping of the electron distribution in the emitting semiconductor band states onto the vacuum states using the appropriate representation of the electron flux **j** over the photoemission barrier; namely, $j \propto p_0 cos\theta$, where $p_0$ is the electron momentum in the vacuum and θ is its angle with respect to the photocathode surface normal. The representation of the electron flux in this manner is forced by the momentum resonance which sets the wave function probability of transmission over the photoemission barrier to unity. Under the assumption of a non-degenerate thermalized electron distribution at temperature $T_e$ in the photocathode band states, the direct mapping for ħΩ → 0 generates the following results [47]:

$$MTE = k_B T_e \left[\frac{3k_B T_e + \chi}{2k_B T_e + \chi}\right] \quad \text{(A1)}$$

$$QE \propto (k_B T_e)^2 [2k_B T_e + \chi]\, exp\left[-\frac{\chi}{k_B T_e}\right] \quad \text{(A2)}$$

for positive electron affinity ($\chi > 0$), and

$$MTE = \frac{|\chi|}{2} + k_B T_e \left[\frac{3k_B T_e + |\chi|}{2k_B T_e + |\chi|}\right] \quad \text{(A3)}$$

$$QE \propto (k_B T_e)^2 [2k_B T_e + |\chi|] \quad \text{(A4)}$$

for negative electron affinity ($\chi < 0$), where $k_B$ is Boltzmann's constant. These relations also assume a low QE (less than about 10%) such that the emission is a small perturbation on the electron distribution in the semiconductor band states (i.e. produces a negligible change in $T_e$).

For photocathode materials like diamond, which have a large phonon energy ħΩ and a large intrinsic Fröhlich coupling constant $g_0$, equations A1-A4 describing the MTE and QE resulting from the FC emission process must be modified to account for energy shifts and multi-phonon emission. First, the real part of the electron self-energy shift associated with the polaron quasi-particle [39,70] produces an effective electron affinity $\chi + g\hbar\Omega$, where $\chi$ is the electron affinity associated with the actual band minimum (calculated by, for example, density functional theory (DFT)) and $g = g_0\sqrt{m^*/m_0}$ with $m^*$ the effective electron mass of the band in the emission (transport) direction [47]. Second, the electrons in the thermalized polaron quasi-particle distribution can either be directly FC emitted ($n = 0$) or be emitted through (multi)-phonon emission ($n \geq 1$) [39] with the strength of the $n^{\text{th}}$ FC emission process given by Poisson statistics as $e^{-g}g^n/n!$ [39,73]. For the case when $k_B T_e$ is significantly less than $\hbar\Omega$, this generates a cascade of emission 'replicas' of the portions of the thermalized electron distribution that can emit (i.e., are above the vacuum level after losing $n\hbar\Omega$ of energy) – as observed by Rameau et al. [39]. In other words, in the above expressions for FC emission, $\chi \to \chi + (g - n)\hbar\Omega$ for $n = 0, 1, 2, 3 \ldots$ with the $n^{\text{th}}$ emission weighted by $e^{-g}g^n/n!$, and one needs to sum over $n$ to determine the total electron emission so that, for example, equation A4 for the total QE for NEA is modified to read

$$QE \propto \sum_{n=0}^{N} \frac{e^g g^n}{n!} (k_B T_e)^2 [2k_B T_e + |\chi + (g+n)\hbar\Omega|] \qquad \text{(A5)}$$

where the upper limit $N$ in the summation is determined by physical constraints (e.g., lack of vacuum states) or practical expediency associated with the rapid convergence of the series.

Applying this procedure to evaluate the total MTE is straightforward since, for independent emission processes, it is simply expressed as the sum of the individual MTEs of the $n^{\text{th}}$ emission weighted by the QE of that emission all divided by the total QE. For the case of positive electron affinity, the result is

$$MTE = k_B T_e \frac{\sum_{n=0}^{\infty} \frac{g^n}{n!} [3k_B T_e + \chi + (g+n)\hbar\Omega].exp\left(-\frac{n\hbar\Omega}{k_B T_e}\right)}{\sum_{n=0}^{\infty} \frac{g^n}{n!} [2k_B T_e + \chi + (g+n)\hbar\Omega].exp\left(-\frac{n\hbar\Omega}{k_B T_e}\right)} \qquad \text{(A6)}$$

which is technically valid for $\chi > -g\hbar\Omega$ to ensure that the effective electron affinity $\chi + (n+g)\hbar\Omega$ is positive for $n = 0$; that is, the effective conduction band minimum, including the real part of the electron self-energy shift associated with the polaron quasi-particle [39,72], is below the vacuum level. In practice, the summations need not be extended to infinity as the $g^n/n!$ factor from the Poisson statistics of phonon emission [39,73] ensures that both series quickly converge for sufficiently large $n$. In fact, for diamond, where $g \approx 1$ [39], even without the $\exp[-n\hbar\Omega/(k_B T_e)]$ factor, four phonon emission ($n = 4$) is only ~4% of the peak emission at $n = 1$ and accounts for less than 2% of the total emission. Consequently, termination of the two series at $n = 4$ introduces an uncertainty of significantly less than 1% in the theoretical evaluation of the MTE.

For the case of negative electron affinity, the result is

$$MTE = \frac{\sum_{n=0}^{N} \frac{g^n}{n!} [|\chi + (g+n)\hbar\Omega|^2 + 4k_B T_e|\chi + (g+n)\hbar\Omega| + 6(k_B T_e)^2]}{2\sum_{n=0}^{N} \frac{g^n}{n!} [2k_B T_e + |\chi + (g+n)\hbar\Omega|]} \qquad \text{(A7)}$$

which is valid for a negative effective electron affinity $\chi + (n+g)\hbar\Omega$; that is, for $n$ less than $|\chi|/(\hbar\Omega) - g$ which gives the maximum integer $N$ for the summations. For $n > |\chi|/(\hbar\Omega) - g$, the effective electron affinity for the FC emission process becomes positive and the emission efficiency is strongly reduced due to the $\exp[-n\hbar\Omega/(k_B T_e)]$ factor (equation A6). Consequently, for nearly all cases, the evaluation of the MTE up to $n = N$ is sufficient, even discounting the $g^n/n!$ factor from the Poisson statistics of phonon emission [39,73].

We also note only optical phonon emission is considered in the presented FC emission theory. For the case of a non-degenerate thermalized electron distribution in diamond presented in this article, where $k_B T_e \approx 30$meV and $\hbar\Omega \approx 160$meV [39,66], this is a very good approximation since the expected ratio of phonon emission to phonon absorption is greater than 100:1 [61]. Only when $k_B T_e$ is greater than $\hbar\Omega$ will this approximation start to break down.

*Contact author: andreas@uic.edu

*Contact author: andreas@uic.edu

*Contact author: andreas@uic.edu

*Contact author: andreas@uic.edu

*Contact author: andreas@uic.edu

*Contact author: andreas@uic.edu